\documentstyle[12pt]{article}
\newcommand{\nc}{\newcommand}
\nc{\renc}{\renewcommand}

\nc{\etal}{\mbox{\it et al. }}
\nc{\ie}{{\it i.e.}}
\nc{\eg}{{\it e.g.}}

\renc{\thefootnote}{\arabic{footnote}}
\nc{\capt}[1]{{\bf Figure.} {\small\sl #1}}

\nc{\eqs}[2]{\mbox{Eqs.~(\ref{#1},\,\ref{#2})}}
\nc{\eq}[1]{\mbox{Eq.~(\ref{#1})}}

\nc{\figs}[2]{\mbox{Figs.~(\ref{#1},\,\ref{#2})}}
\nc{\fig}[1]{\mbox{Fig~.(\ref{#1})}}

\nc{\tag}[1]{\label{#1} \marginpar{{\footnotesize #1}}}
\nc{\mtag}[1]{\label{#1} \mbox{\marginpar{{\footnotesize #1}}}}
\renc{\baselinestretch}{1.2}
\newlength{\overeqskip}
\newlength{\undereqskip}
\nc{\be}[1]{\begin{equation} \mbox{$\label{#1}$}}
\nc{\bea}[1]{\begin{eqnarray} \mbox{$\label{#1}$}}
\nc{\Section}[2]{\section{#2}\label{#1}}
\nc{\Bibitem}[1]{\bibitem{#1}}
\nc{\Label}[1]{\label{#1}}

\nc{\eea}{\vspace{\undereqskip}\end{eqnarray}}
\nc{\ee}{\vspace{\undereqskip}\end{equation}}
\nc{\bdm}{\begin{displaymath}}
\nc{\edm}{\end{displaymath}}
\nc{\dpsty}{\displaystyle}
\nc{\bc}{\begin{center}}
\nc{\ec}{\end{center}}
\nc{\ba}{\begin{array}}
\nc{\ea}{\end{array}}
\nc{\bab}{\begin{abstract}}
\nc{\eab}{\end{abstract}}
\nc{\btab}{\begin{tabular}}
\nc{\etab}{\end{tabular}}
\nc{\bit}{\begin{itemize}}
\nc{\eit}{\end{itemize}}
\nc{\ben}{\begin{enumerate}}
\nc{\een}{\end{enumerate}}
\nc{\bfig}{\begin{figure}}
\nc{\efig}{\end{figure}}
\nc{\arreq}{&\!=\!&}
\nc{\arrmi}{&\!-\!&}
\nc{\arrpl}{&\!+\!&}
\nc{\arrap}{&\!\!\!\approx\!\!\!&}
\nc{\non}{\nonumber\\*}
\nc{\align}{\!\!\!\!\!\!\!\!&&}

\def\lsim{\; \raise0.3ex\hbox{$<$\kern-0.75em
      \raise-1.1ex\hbox{$\sim$}}\; }
\def\gsim{\; \raise0.3ex\hbox{$>$\kern-0.75em
      \raise-1.1ex\hbox{$\sim$}}\; }
\nc{\DOT}{\hspace{-0.08in}{\bf .}\hspace{0.1in}}
\nc{\Laada}{\hbox {$\sqcap$ \kern -1em $\sqcup$}}
\nc\loota{{\scriptstyle\sqcap\kern-0.55em\hbox{$\scriptstyle\sqcup$}}}
\nc\Loota{{\sqcap\kern-0.65em\hbox{$\sqcup$}}}
\nc\laada{\Loota}
\nc{\qed}{\hskip 3em \hbox{\BOX} \vskip 2ex}

\nc{\real}{{\rm I \! R}}
\nc{\Z}{{\sf Z \!\!\! Z}}
\nc{\complex}{{\rm C\!\!\! {\sf I}\,\,}}
\def\bigid{\leavevmode\hbox{\small1\kern-3.8pt\normalsize1}}
\def\id{\leavevmode\hbox{\small1\kern-3.3pt\normalsize1}}
\nc{\slask}{\!\!\!/}
\nc{\bis}{{\prime\prime}}
\nc{\pa}{\partial}
\nc{\na}{\nabla}
\nc{\ra}{\rangle}
\nc{\la}{\langle}
\nc{\goto}{\rightarrow}
\nc{\swap}{\leftrightarrow}

\nc{\EE}[1]{ \mbox{$\cdot10^{#1}$} }
\nc{\abs}[1]{\left|#1\right|}
\nc{\at}[2]{\left.#1\right|_{#2}}
\nc{\norm}[1]{\|#1\|}
\nc{\abscut}[2]{\Abs{#1}_{\scriptscriptstyle#2}}
\nc{\vek}[1]{{\rm\bf #1}}
\nc{\integral}[2]{\int\limits_{#1}^{#2}}
\nc{\inv}[1]{\frac{1}{#1}}
\nc{\dd}[2]{{{\partial #1}\over{\partial #2}}}
\nc{\ddd}[2]{{{{\partial}^2 #1}\over{\partial {#2}^2}}}
\nc{\dddd}[3]{{{{\partial}^2 #1}\over
	{\partial #2 \partial #3}}}
\nc{\dder}[2]{{{d #1}\over{d #2}}}
\nc{\ddder}[2]{{{d^2 #1}\over{d {#2}^2}}}
\nc{\dddder}[3]{{d^2 #1}\over
	{d #2 d #3}}
\nc{\dx}[1]{d\,^{#1}x}
\nc{\dy}[1]{d\,^{#1}y}
\nc{\dz}[1]{d\,^{#1}z}
\nc{\dl}[1]{\frac{d\,^{#1}l}{(2\pi)^{#1}}}
\nc{\dk}[1]{\frac{d\,^{#1}k}{(2\pi)^{#1}}}
\nc{\dq}[1]{\frac{d\,^{#1}q}{(2\pi)^{#1}}}

\nc{\cc}{\mbox{$c.c.$ }}
\nc{\hc}{\mbox{$h.c.$ }}
\nc{\cf}{cf.\ }
\nc{\erfc}{{\rm erfc}}
\nc{\Tr}{{\rm Tr\,}}
\nc{\tr}{{\rm tr\,}}
\nc{\pol}{{\rm pol}}
\nc{\sign}{{\rm sign}}
\nc{\bfT}{{\bf T }}

\nc{\cA}{{\cal A}}
\nc{\cB}{{\cal B}}
\nc{\cD}{{\cal D}}
\nc{\cE}{{\cal E}}
\nc{\cG}{{\cal G}}
\nc{\cH}{{\cal H}}
\nc{\cL}{{\cal L}}
\nc{\cO}{{\cal O}}
\nc{\cT}{{\cal T}}
\nc{\cN}{{\cal N}}
\nc{\rvac}[1]{|{\cal O}#1\rangle}
\nc{\lvac}[1]{\langle{\cal O}#1|}
\nc{\rvacb}[1]{|{\cal O}_\beta #1\rangle}
\nc{\lvacb}[1]{\langle{\cal O}_\beta #1 |}
\nc{\bb}{\bar{\beta}}
\nc{\bt}{\tilde{\beta}}
\nc{\ctH}{\tilde{\cal H}}
\nc{\chH}{\hat{\cal H}}
\nc{\1}{\aa}
\nc{\2}{\"{a}}
\nc{\3}{\"{o}}
\nc{\4}{\AA}
\nc{\5}{\"{A}}
\nc{\6}{\"{O}}
\nc{\al}{\alpha}
\nc{\g}{\gamma}
\nc{\Del}{\Delta}
\nc{\e}{\epsilon}
\nc{\eps}{\epsilon}
\nc{\lam}{\lambda}
\nc{\om}{\omega}
\nc{\Om}{\Omega}
\nc{\ve}{\varepsilon}
\nc{\mn}{{\mu\nu}}
\nc{\k}{\kappa}
\nc{\vp}{\varphi}

\nc{\advp}[3]{{\it  Adv.\ in\ Phys.\ }{{\bf #1} {(#2)} {#3}}}
\nc{\annp}[3]{{\it  Ann.\ Phys.\ (N.Y.)\ }{{\bf #1} {(#2)} {#3}}}
\nc{\apl}[3]{{\it  Appl. Phys. Lett. }{{\bf #1} {(#2)} {#3}}}
\nc{\apj}[3]{{\it  Ap.\ J.\ }{{\bf #1} {(#2)} {#3}}}
\nc{\apjl}[3]{{\it  Ap.\ J.\ Lett.\ }{{\bf #1} {(#2)} {#3}}}
\nc{\app}[3]{{\it Astropart.\ Phys.\ }{{\bf #1} {(#2)} {#3}}}  
\nc{\cmp}[3]{{\it  Comm.\ Math.\ Phys.\ }{{ \bf #1} {(#2)} {#3}}}
\nc{\cqg}[3]{{\it  Class.\ Quant.\ Grav.\ }{{\bf #1} {(#2)} {#3}}}
\nc{\epl}[3]{{\it  Europhys.\ Lett.\ }{{\bf #1} {(#2)} {#3}}}
\nc{\ijmp}[3]{{\it Int.\ J.\ Mod.\ Phys.\ }{{\bf #1} {(#2)} {#3}}}
\nc{\ijtp}[3]{{\it Int.\ J.\ Theor.\ Phys.\ }{{\bf #1} {(#2)} {#3}}}
\nc{\jmp}[3]{{\it  J.\ Math.\ Phys.\ }{{ \bf #1} {(#2)} {#3}}}
\nc{\jpa}[3]{{\it  J.\ Phys.\ A\ }{{\bf #1} {(#2)} {#3}}}
\nc{\jpc}[3]{{\it  J.\ Phys.\ C\ }{{\bf #1} {(#2)} {#3}}}
\nc{\jap}[3]{{\it J.\ Appl.\ Phys.\ }{{\bf #1} {(#2)} {#3}}}
\nc{\jpsj}[3]{{\it J.\ Phys.\ Soc.\ Japan\ }{{\bf #1} {(#2)} {#3}}}
\nc{\lmp}[3]{{\it Lett.\ Math.\ Phys.\ }{{\bf #1} {(#2)} {#3}}}
\nc{\mpl}[3]{{\it  Mod.\ Phys.\ Lett.\ }{{\bf #1} {(#2)} {#3}}}
\nc{\ncim}[3]{{\it  Nuov.\ Cim.\ }{{\bf #1} {(#2)} {#3}}}
\nc{\np}[3]{{\it  Nucl.\ Phys.\ }{{\bf #1} {(#2)} {#3}}}
\nc{\pr}[3]{{\it Phys.\ Rev.\ }{{\bf #1} {(#2)} {#3}}}
\nc{\pra}[3]{{\it  Phys.\ Rev.\ A\ }{{\bf #1} {(#2)} {#3}}}
\nc{\prb}[3]{{\it  Phys.\ Rev.\ B\ }{{{\bf #1} {(#2)} {#3}}}}
\nc{\prc}[3]{{\it  Phys.\ Rev.\ C\ }{{\bf #1} {(#2)} {#3}}}
\nc{\prd}[3]{{\it  Phys.\ Rev.\ D\ }{{\bf #1} {(#2)} {#3}}}
\nc{\prl}[3]{{\it Phys\ Rev.\ Lett.\ }{{\bf #1} {(#2)} {#3}}}
\nc{\pl}[3]{{\it  Phys.\ Lett.\ }{{\bf #1} {(#2)} {#3}}}
\nc{\prep}[3]{{\it Phys\. Rep.\ }{{\bf #1} {(#2)} {#3}}}
\nc{\prsl}[3]{{\it Proc.\ R.\ Soc.\ London\ }{{\bf #1} {(#2)} {#3}}}
\nc{\ptp}[3]{{\it  Prog.\ Theor.\ Phys.\ }{{\bf #1} {(#2)} {#3}}}
\nc{\ptps}[3]{{\it  Prog\ Theor.\ Phys.\ suppl.\ }{{\bf #1} {(#2)} {#3}}}
\nc{\physa}[3]{{\it  Physica\ A\ }{{\bf #1} {(#2)} {#3}}}
\nc{\physb}[3]{{\it  Physica\ B\ }{{\bf #1} {(#2)} {#3}}}
\nc{\phys}[3]{{\it Physica\ }{{\bf #1} {(#2)} {#3}}}
\nc{\rmp}[3]{{\it  Rev.\ Mod.\ Phys.\ }{{\bf #1} {(#2)} {#3}}}
\nc{\rpp}[3]{{\it Rep.\ Prog.\ Phys.\ }{{\bf #1} {(#2)} {#3}}}
\nc{\sjnp}[3]{{\it Sov.\ J.\ Nucl.\ Phys.\ }{{\bf #1} {(#2)} {#3}}}
\nc{\spjetp}[3]{{\it Sov.\ Phys.\ JETP\ }{{\bf #1} {(#2)} {#3}}}
\nc{\yf}[3]{{\it Yad.\ Fiz.\ }{{\bf #1} {(#2)} {#3}}}
\nc{\zetp}[3]{{\it Zh.\ Eksp.\ Teor.\ Fiz.\  }{{\bf #1}  {(#2)} {#3}}}
\nc{\zp}[3]{{\it Z.\ Phys.\ }{{\bf #1} {(#2)} {#3}}}
\nc{\ibid}[3]{{\sl ibid.\ }{{\bf #1} {#2} {#3}}}
\nc{\rf}[1]{(\ref{#1})}
\nc{\nn}{\nonumber \\*}
\nc{\bfB}{\bf{B}}
\nc{\bfv}{\bf{v}}
\nc{\bfx}{\bf{x}}
\nc{\bfy}{\bf{y}}
\nc{\vx}{\vec{x}}
\nc{\vy}{\vec{y}}
\nc{\oB}{\overline{B}}
\nc{\oI}{\overline{I}}
\nc{\oR}{\overline{R}}
\nc{\rar}{\rightarrow}
\nc{\ti}{\times}
\nc{\slsh}{\hskip-5pt/}
\nc{\sm}{Standard~Model~}
\nc{\MP}{M_{\rm Pl}}
\nc{\tp}{t_{\rm Pl}}
\nc{\ave}{\bar{E}}

\renc{\min}{p_{\rm min}}
\renc{\max}{p_{\rm max}}
\nc{\pmin}{p_{\rm min}}
\nc{\pmax}{p_{\rm max}}
\nc{\fo}{f_0}
\nc{\foi}{f_{0,i}\,}
\nc{\fop}{f_0^P}
\nc{\fou}{f_0^U}
\def\sepand{\rule{14cm}{0pt}\and}
\nc{\eff}{{\rm eff}}
\nc{\MT}{M_{\rm T}}
\nc{\ML}{M_{\rm L}}
\nc{\kk}{\vek{k}}
\nc{\pp}{{\rm p}}
\nc{\cb}{critical bubble~}
\nc{\cbs}{critical bubbles~}
\nc{\scb}{subcritical bubble~}
\nc{\scbs}{subcritical bubbles~}
\begin{document}

{\title{{\hfill {{\small  TURKU-FL-P26-97 
        }}\vskip 1truecm}
{\bf CP-Violation and Baryogenesis in The Low Energy Minimal 
Supersymmetric Standard Model}}

 
\author{
{\sc Tuomas Multam\" aki$^{1}$}\\
{\sl and}\\
{\sc Iiro Vilja$^{2}$ }\\ 
{\sl Department of Physics,
University of Turku} \\
{\sl FIN-20500 Turku, Finland} \\
\sepand
}
\maketitle}
\vspace{2cm}
\begin{abstract}
\noindent In the context of the minimal supersymmetric extension of the 
Standard Model the effect of a realistic wall profile is studied. It has been 
recently showed that in the presence of light stops the electroweak scale 
phase transition can be strong enough for baryogenesis. In the presence of
non-trivial CP-violating phases of left-handed mixing terms and Higgsino
mass, the largest $n_B/s$ is created when Higgsino and gaugino mass parameters 
are degenerate, $\mu = M_2$. In the present paper we show that realistic wall 
profiles suppress the generated baryon number of the universe, so that quite a
stringent bound $|\sin\phi_\mu | \gsim 0.2$ for $\mu$-phase $\phi_\mu$ can be
inferred.
\end{abstract}
\vfill
\footnoterule
{\small$^1$tuomul@newton.tfy.utu.fi,  $^2$vilja@newton.tfy.utu.fi}
\thispagestyle{empty}
\newpage
\setcounter{page}{1}

The Minimal Supersymmetric Standard Model (MSSM) has appeared to be one of the
most promising candidates to explain the observed baryon asymmetry of the 
universe $n_B/s \sim 10^{-10}$ generated at electroweak scale.
\cite{rev}. Although all requirements are included already in the
Standard Model \cite{rev,Sakharov}, the phase transition \cite{Shapo}
has appeared to be too
weakly first order to preserve the generated baryon asymmetry \cite{many}. 
Also it has been shown that the CP-violation needed for baryogenesis
is too small in the Standard Model \cite{CPSM}.
Therefore some new physics besides the Standard Model is necessarily
needed, provided that the baryon asymmetry is generated during the electroweak 
phase transition. 

Because MSSM is one of the most appealing extensions of the Standard Model,
it has been worthwhile to study whether it is possible to generate and
preserve the baryon asymmetry in it. Indeed, recent analyses show that there
exists a region of the parameter space where the phase transition is strong 
enough \cite{strength}. It is required that $\tan\beta < 3$, the lightest stop
is lighter than the top quark and the lightest Higgs must be detectable by LEP2.
The bound given above may relax due to two and higher -loop effects, which
seem to strengthen the phase transition \cite{two}. Unlike the Standard Model
where the source of CP-violation is solely the Cabibbo-Kobayashi-Maskawa
matrix, MSSM contains an additional source due to the soft supersymmetry 
breaking parameters which are related to the stop mixing angle.

In a recent paper Carena {\it et al.} \cite{CQRVW} has analysed the region of 
supersymmetric parameter space where the baryon asymmetry generated at the
electroweak scale is consistent with the 
observed one. The generation and survival of large enough $n_B/s \simeq 
4 \times 10^{-11}$ seem to require that $M_Z \lsim m_{\tilde t} \lsim m_t$,
the mass of the lightest Higgs boson is bounded by $m_H < 80$ GeV whereas 
CP-odd boson has mass $m_A \gsim 150$ GeV. The analysis gave dependence of
$n_B/s$ on Higgsino mass parameters $|\mu |$ and its phase $\phi_\mu$ 
as well as gaugino mass parameters $M_1$ and $M_2$.
The optimal choice of parameters showed up to be $|\mu | \simeq M_2$, 
$|\sin \phi_\mu | \gsim 0.06$, and
the dependence on $M_1$ is weak, so that it can be chosen to be equal to $M_2$.
In the analysis of Carena {\it et al.} it was chosen the left-handed stop mass
parameter $m_Q$ to be $m_Q = 500$ GeV, effective soft supersymmetry breaking 
parameter $\tilde A_t = 0$ and right-handed stop
mass parameter $m_U = - \tilde m_U < 0$
\be{mUcrit}
\tilde m_U \lsim m_U^{{\rm crit}} \equiv \left ( {m_H^2 v_0^2 g_3^2\over 12}
\right )^{1/4}
\ee
to obtain the most optimistic bounds (i.e. maximize $n_B/s$) 
\footnote{About restrictions and validity of these results, see \cite{CQRVW}.}.
With these parameter values the CP-violating source is generated essentially by
Higgsino and gaugino currents and the right-handed stop contribution is
negligible, so that without a loss of generality one can set 
$\sin (\phi_\mu + \phi_A) = 0$.
The bound (\ref{mUcrit}) is due to the colour non-breaking condition, {\it i.e} 
no colour breaking minimum must be deeper than the normal electroweak 
breaking (and colour conserving) minimum.
It is defined at zero-temperature, thus $v_0 = 246.22$ GeV. Also it was used 
$v_w = 0.1$ and $L_w = 25/T$. With these conditions a large enough baryon 
asymmetry could have been created.

In the paper \cite{CQRVW} the baryon asymmetry was inferred by first
calculating the CP-violating sources and then solving the relevant Boltzmann
equations. It was shown that the baryon to entropy ratio reads
\be{bno}
{n_B\over s} = - g(k_i) {{\cal A} \bar D \Gamma_{ws}\over v_w^2 s}, 
\ee
where $g(k_i)$ is a numerical coefficient depending on the degrees of freedom,
$\bar D$ the effective diffusion rate, 
$\Gamma_{ws} = 6 \kappa \alpha_w^4 T$ ($\kappa = 1$) \cite{AK} the weak 
sphaleron rate, $v_w$ the wall velocity. The entropy density $s$ is given by 
\be{entropy}
s = {2 \pi^2 g_{*s} T^3 \over 45},
\ee
where $g_{*s}$ is the effective number of relativistic degrees of freedom. The
coefficient ${\cal A}$  was shown to be a certain integral over the source
$\tilde \gamma(u) = v_w f(k_i) \partial_u J^0(u)$:
\be{A}
{\cal A} = {1\over \bar D \lambda_+} \int_0^\infty du \tilde \gamma (u) 
e^{-\lambda_+ u},
\ee
where $\lambda_+ = (v_w + \sqrt{v_w^2 + 4 \tilde \Gamma \bar D})/(2\bar D)$ and
the wall was defined so that it begins at $u = 0$. Here $u$ is the co-moving 
coordinate $u = z + v_w t$ supposing that the wall moves in the direction of
z-axis. ($f(k_i)$ is again a 
coefficient depending on the number of degrees of freedom present in thermal 
path and related to the definition of the effective source \cite{CQRVW,Nelson}.)
Thus the coefficient ${\cal A}$ is dependent on the actual wall shape via
\be{integral}
{\cal A} \propto I \equiv \int_0^{\infty} du {\partial \over \partial u} 
(H(u)^2{\partial \beta(u) \over \partial u}) e^{- \lambda_+ u},
\ee
where $H = \sqrt{H_1^2 + H_2^2}$, $\tan\beta = H_1/H_2$ and $H_i$'s are
the real parts of the neutral components of the Higgs doublets. 
In \cite{CQRVW} the wall shape was, however, taken {\it ad hoc}. 
It was assumed to have a simple sinusoidal form so that the field $H(u)$ 
can be given by
\be{apwallH}
H_s(u) = {v \over 2}\left [1 - \cos\left ({u\pi\over L_w} \right )\right ]
[\theta(u) - \theta(u - L_w)] + v \theta(u - L_w)
\ee
and the field angle $\beta(u)$ by
\be{apwallbeta}
\beta_s(u) = {\Delta\beta\over 2} \left [1 - \cos\left ({u\pi\over L_w} 
\right )\right ][\theta(u) - \theta(u - L_w)] + \Delta\beta \theta(u - L_w),
\ee
where $\Delta \beta$ is given by $\Delta\beta = \beta(T_0) - 
\arctan(m_1(T_0)/m_2(T_0))$, calculated at the temperature where curvature
of the one-loop effective potential vanishes at the origin. Inserting these 
{\it Anz\" atse} to Eq. (\ref{integral}), we obtain a corresponding contribution
to the CP-violation
\be{int2}
I_s = \int_0^{\infty} du {\partial\over \partial u}(H_s(u)^2{\partial\beta_s(u) 
\over \partial u}) e^{- \lambda_+ u}.
\ee
Using this approximation, the results of \cite{CQRVW} was inferred.

In the present paper a more realistic prescription of wall shape is used. Working
at the critical temperature $T_c$ and using the one-loop resummed effective
potential \cite{potential} we numerically find the path of smallest gradient 
$\gamma_g$
from  $(H_1, H_2) = (0, 0)$ to $(H_1, H_2) = (v_1, v_2) \equiv 
(v_1(T_c), v_2(T_c))$, which well 
approximates the true solution. Moreover, using path $\gamma_g$,
a upper bound for $I$ is necessarily obtained. The true solution lies 
necessarily between $\gamma_g$
and straight line from $(0, 0)$ to $(v_1, v_2)$ (which leads to $n_B = 0$)
as can be concluded by studying the Lagrangean\footnote{This is true providing
that the field is not strongly oscillating within the wall, but is a
smooth configuration.}. Thus $|I|$ along such a path is smaller than along $\gamma_g$. 

The effective potential for MSSM at finite temperature can be 
expressed in three parts \cite{potential}
\be{pot}
V_{eff}(\phi,T) = V_0(\phi ) + V_1(\phi ) + V_{1, T}(\phi ) +
\Delta V_T(\phi ),
\ee
where $V_0$ is the tree level zero-temperature potential, $V_1$ the 
renormalized 1-loop zero-temperature potential, $V_{1, T}$ the 1-loop finite
temperature potential and $\Delta V_T$ the daisy-resummed part. They are 
given by
\be{part1}
V_0(H) = m_1^2 H_1^2 + m_2^2 H_2^2 + 2 m_{12}^2 H_1 H_2 + 
{g^2 + g'^2\over 8} (H_1^2 - H_2^2)^2,
\ee
\be{part2}
V_1(H) = \sum_{t,\tilde{t}_{1,2},W,Z} {n_i\over 64\pi^2} 
m_i^4(H) (\ln{m_i^2(H)\over m_Z^2} - {3\over 2}), 
\ee
\be{part3}
V_{1, T} (H) = {T^4\over 2\pi^2} \sum_{t,\tilde{t}_{1,2},W,Z} n_i 
J_i[{m_i^2(H)\over T^2}]
\ee
and
\be{part4}
\Delta V_T(H) = -{T\over 12\pi} \sum_i n_i[\bar{m_i^3(H,T)} - m_i^3(H)], 
\ee
where $m_i(H)$ and $m_i(H, T)$ are the zero temperature and temperature
corrected field dependent masses, respectively, $n_i$ are the degrees of 
freedom of each particle (including $-$ -sign for fermions) and 
\be{J}
J_i(x^2) = \int dy\, y^2 \ln ( 1 \pm e^{- \sqrt{y^2 + x^2}}).
\ee 
Note, that we have neglected the b-quark, $\tilde {\rm b}$-squark as wall as
other generation contributions as small ones. The heavy supersymmetric
particles do not contribute neither.
The mass parameters $m_1,\ m_2,\ m_{12}$, are related to $\beta$, $m_A$,
$m_H$ and other parameters of the theory, as given in \cite{potential}. 
Using these formulas, we can solve $\gamma_g$ and define the corresponding 
CP-violation integral 
\be{int1}
I_\gamma = \int_0^{\infty} du {\partial\over \partial u}(H(u)^2{\partial \beta(u) \over \partial u}) e^{- \lambda_+ u}
\ee
at the critical temperature along the path $\gamma_g$. It also shows up
that the form of profiles is, with good acccuracy, the form of a kink. Indeed,
if we reparametrise the field $(H_1, H_2)$ to a component pointing towards 
$(v_1,\ v_2)$, $H_\parallel $ and to a component orthogonal to that,
$H_\perp $, it appears that the ratio of maximum value of $H_\perp$ to
$v = \sqrt{v_1^2 + v_2^2}$ is in any case smaller than 0.004. Thus the bending 
of the path, {\it I.e.} the
deviation from a straight line is small. (Note, that here $v$ is the value of
vacuum at $T_c$, thus not equal to $v_0 = 247$ GeV.)

The ratio $I_\gamma /I_s$ gives immediately the supression of $n_B/s$ with 
respect to the results of Carena {\it et al.} \cite{CQRVW}. 
Hence the value needed for
soft supersymmetry mixing phase $\sin\phi_\mu$ is increased by factor
$I_\gamma /I_s$. In the Figure 1. we have presented the value of the integral $I_s$
as a function of $\mu$ for several values of $m_A$. From the figure it can be 
read out that increasing $\mu$ decreases the (absolute) value of $I_\gamma$ so
that for $\mu > 250$ GeV inequality $ |I_\gamma| < 1$ holds. Also it can be found that increasing
$m_A$ with factor $r$ decreases $I_\gamma$ roughly by factor $1/r$ within 
the parameter 
range studied. This behaviour is likely be  more general than just restricted
to analysis of \cite{CQRVW}, because the amount of CP-violation is in general
proportional to the change of $\beta$ on the path from the origin to the 
non-trivial vacuum. 

In Figure 2. a comparison to the result of Carena {\it et al.} \cite{CQRVW}
is made. We have plotted the ratio $I_\gamma /I_s$ also as a function of 
$\mu$ with several values of $m_A$. There appears a clear tendency of 
additional 
suppression: for very small $\mu$, $I_\gamma /I_s \simeq 0.4$, whereas for $\mu \simeq
250$ GeV, $I_\gamma /I_s \simeq 0.1$. For optimal values of $\mu$, 
150 GeV $\leq \mu \leq$ 250 GeV the supression factor is at 
least 0.3. Also it appers that the dependence of this suppression factor on 
CP-odd boson mass $m_A$ is relatively weak. Taking in to the account that 
according to the analysis of \cite{CQRVW} it is required that 
$|\sin\phi_\mu| \gsim 0.06$, this is now converted to a more stringent bound
$|\sin\phi_\mu| \gsim 0.2$ which remarkably weakens the possibility of 
baryogenesis in MSSM. However, for $\mu \simeq 250$ GeV the bound rises to
$|\sin\phi_\mu | \gsim 0.6$ which possibly is already too large.

In the present paper we have calculated the amount of CP-violation in the
bubble wall of the minimal supersymmetric extension of the Standard Model at
the electroweak scale phase transition. It has
shown up that the previous estimates tend to be too optimistic and an extra
suppression of (at least) 0.3 is found. This tends to make the electroweak 
baryogenesis in MSSM more difficult and less likely. However, more 
analysis on the model is needed, in particular to clarify how higher
corrections (more scalar insertions) to the CP-violating source behave. If the 
higher corrections to the source expand in the powers of 
$(H_1\partial H_2 - H_2\partial H_1)$ no help from them is expected. If they,
however, order by some other expansion parameter, their contribution to the
source may be remarkably large. Unfortunately this may also lead to the 
situation where non-perturbative effects are important. Also the effect of 
higher order corrected effective potential remains to be studied. The
two-loop corrections tend to strenghten the phase transition and thus relax
the bounds \cite{two}. The two-loop contributions may, however, 
affect also directly the
value of the integral $I$ needed in the calculation of CP-violating source.
On the other hand, also the sphaleron rate $\Gamma_{ws}$ is under discussion
\cite{AJ}
and changes on that may change significantly the conclusions made about
baryogenesis.
\vskip 1truecm

\noindent {\bf Acknowledgements.} The authors thank A. Riotto for 
discussions and J. Sirkka for technical advice.

\newpage

\newpage
%
\noindent {\Large{\bf Figure captions}}
\vskip .5truecm
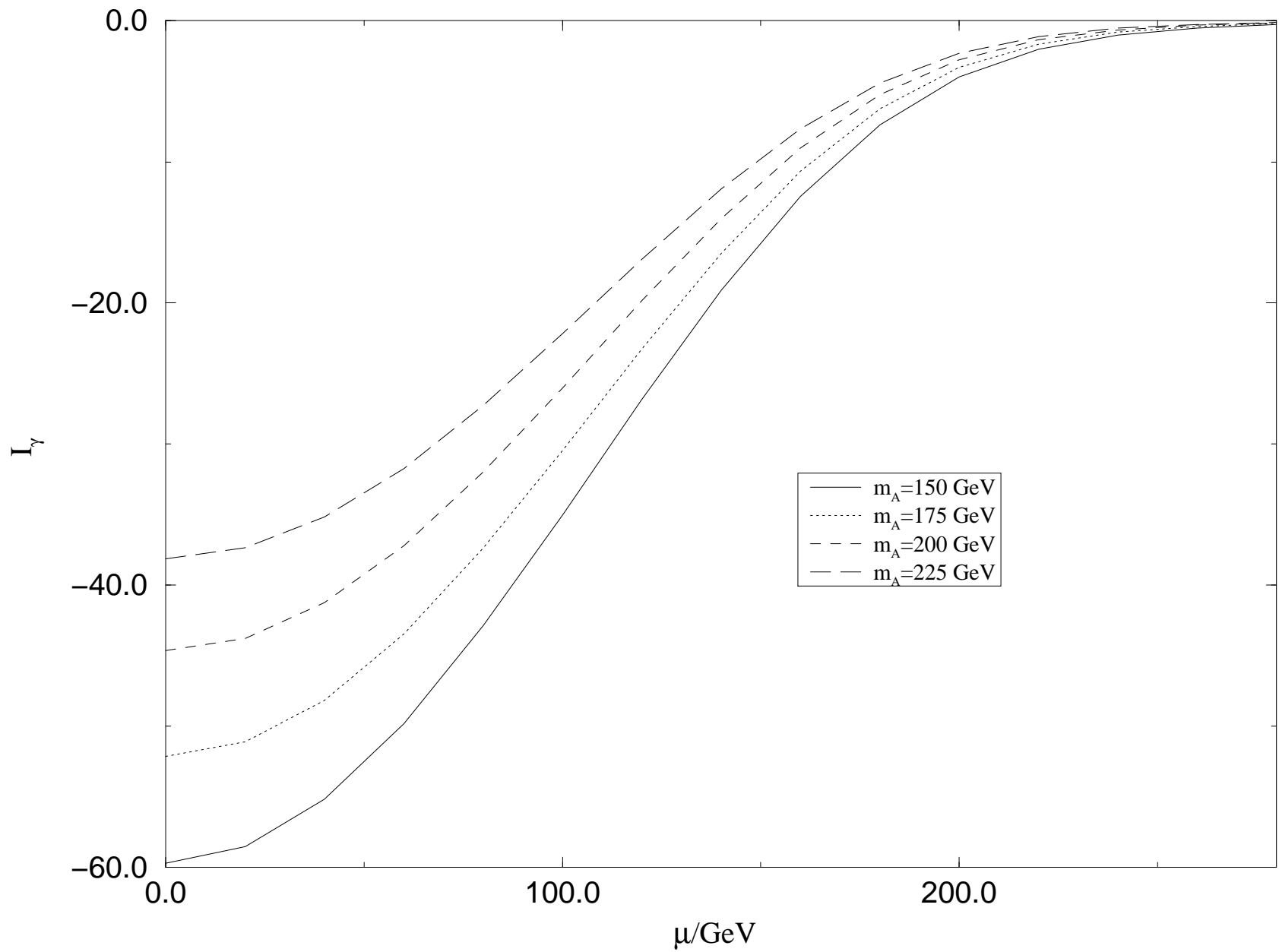
\noindent Figure 1. Values of the integral $I_\gamma =
 \int_0^\infty (H^2\beta')'
e^{-\lambda_+ u}$ (where comma stands for $u$ -derivative) as a function of the soft supersymmetry parameter $\mu = M_1 = M_2$ with $v_w = 0.1$, $m_Q = 500$ GeV, 
$\tilde m_U = \tilde m_U^{{\rm crit}}$, $\tan\beta = 2$ and $\tilde A_t = 0$. 
\vskip .5truecm
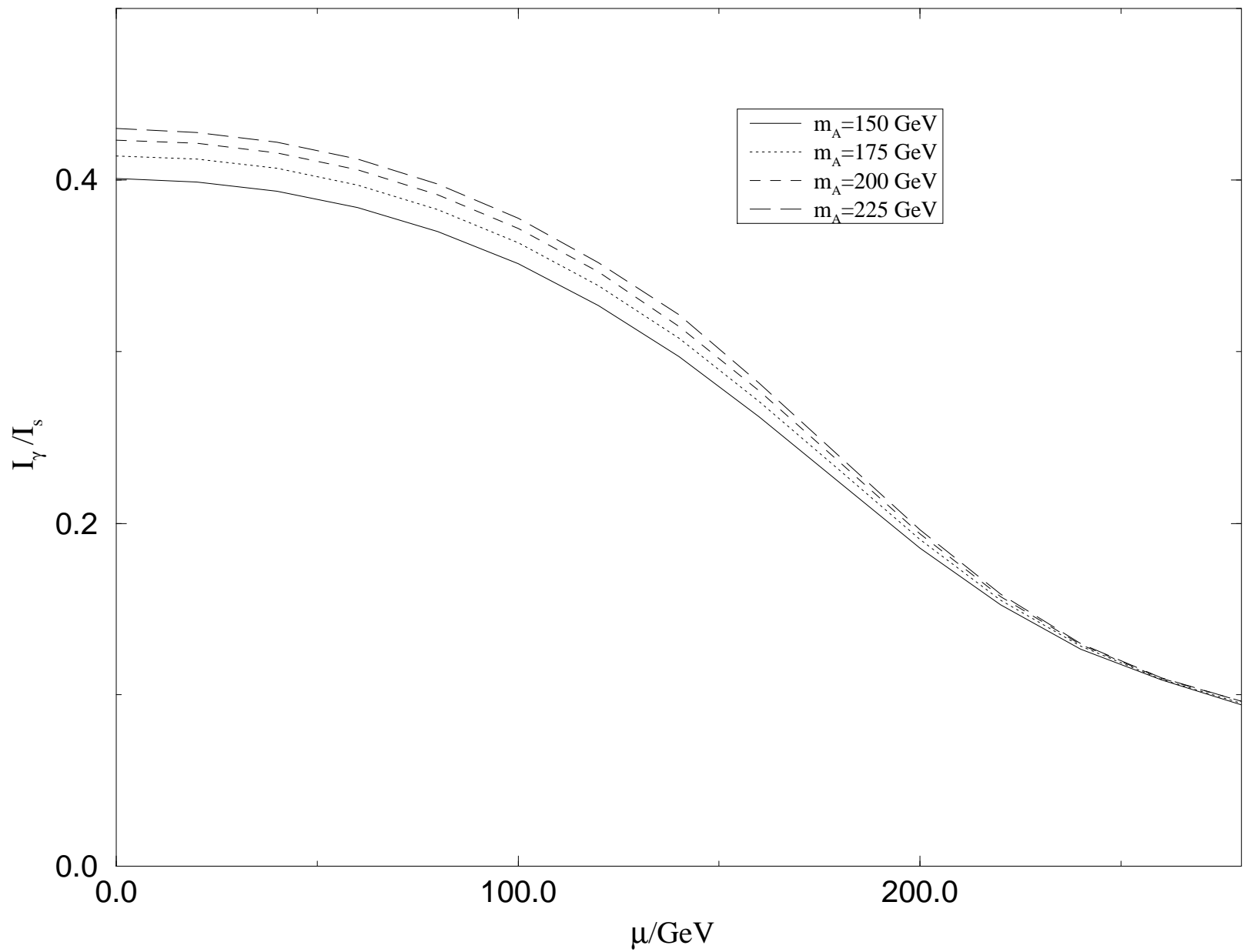
\noindent Figure 2. Values of the ratio $I_\gamma /I_s$ as a function of 
$\mu = M_1 = M_2$ with $v_w = 0.1$, $m_Q = 500$ GeV, $\tilde m_U = 
\tilde m_U^{{\rm crit}}$, $\tan\beta = 2$ and $\tilde A_t = 0$. For $I_s$ wall 
width $L_w = 25/T$ is used.
\end{document}